# Circularly Polarized Resonant Rayleigh Scattering and Skyrmions in the $\nu = 1$ Quantum Hall Ferromagnet

V. Bellani[1], F. Rossella[1], M. Amado[2,3], E. Diez[3], K. Kowalik[4,*], G. Biasiol[5] and L. Sorba[6]
[1]Dipartimento di Fisica "A.Volta" and CNISM, Università degli Studi di Pavia, 27100 Pavia, Italy
[2]GISC, Departamento de Física de Materiales, Universidad Complutense, 28040 Madrid, Spain
[3]Laboratorio de Bajas Temperaturas, Universidad de Salamanca, 37008 Salamanca, Spain
[4]Laboratoire National des Champs Magnétiques Intenses, 38042 Grenoble, France
[5]Istituto Officina dei Materiali CNR-TASC, 34012 Trieste, Italy
[6]NEST, Istituto Nanoscienze-CNR and Scuola Normale Superiore, 56126 Pisa, Italy
(Dated: April 11, 2011)

We use the circularly polarized resonant Rayleigh scattering (RRS) to study the quantum Hall ferromagnet at $\nu = 1$. At this filling factor we observe a right handed copolarized RRS which probes the Skyrmion spin texture of the electrons in the photoexcited grounds state. The resonant scattering is not present in the left handed copolarization, and this can be related to the correlation between Skymionic effects, screening and spin wave excitations. These results evidence that RRS is a valid method for the study of the spin texture of the quantum Hall states.

The interaction between the electrons in the quantum Hall regime gives raise to collective ground states. At filling factor $\nu = 1$ the two dimensional electron system turns into an itinerant ferromagnet while moving away from this filling factor the system depolarizes rapidly due to the formation of a spin texture called Skyrmion.

The Skyrmions have been originally observed using optically pumped nuclear magnetic resonance by Barret et al. [1]. Successively the depolarization of the ferromagnet very close to $\nu = 1$ and the related spin texture have been studied mainly by means of interband optical absorption [2–4] and inelastic light scattering [5]. The temperature affects the ferromagnetic order, and at temperatures larger than the Zeeman gap the formation of low energy spin-excitons separates the ferromagnet into clusters. However the lateral confinement of these clusters pushes the energies of the spin-excitons up, thus preserving the ferromagnetic order locally [6]. Indeed the Skyrmion texture can be observed in optical absorption experiments at liquid helium temperature [2, 3, 18] up to several Kelvin [2, 3].

In this work we evidence the Skyrmion spin texture at $\nu = 1$ by means of elastic resonant Rayleigh scattering (RRS), in a one side modulation doped GaAs-AlGaAs quantum well. In this experiment the incident light is resonant with the photoluminescence (PL) from the recombination between the 2DEG electrons and the photogenerated holes. The selection of the circular polarization of the incident and scattered light allows to probe the different Zeeman split sublevels. Close to $\nu = 1$ we observe a right handed copolarized RRS with the Skyrmion spin texture, while the left handed copolarized spectra does not present such RRS, a behaviour that can be related to the correlation between the Skyrmion, screening and spin waves [7, 8].

RRS proved to be a valid tool to study the quantum mechanics of the electrons and the dephasing process in low dimensional systems [9], and has been used to study the quantum Hall phases in a double coupled 2DEG [10] in order to investigate the formation of a non-uniform electron fluid close to the phase transition towards the compressible state. The light emission from the 2DEG, excited with a laser radiation resonant to the transition energy of the exciton, comprises a contribution which maintains its coherence with the excitation and one which does not. The former is the RRS which depends on eigenenergies and wave-function properties including the spin, while the latter is the (incoherent) luminescence [11, 12].

The experiment was performed on a 20 nm-thick one-side modulation doped GaAs/AlGaAs single quantum well with carrier density $n = 1.8 \times 10^{11}$ cm$^{-2}$ and mobility $\mu = 1.6 \times 10^6$ cm$^{-2}$ V$^{-1}$ s$^{-1}$ measured at 1.4 K. Recent studies show that in 2DEG with intermediate electron density of about $10^{11}$ cm$^{-2}$ and mobility between $1 \times 10^6$ and $3 \times 10^6$ cm$^{-2}$ V$^{-1}$ s$^{-1}$ the interplay between disorder and Coulomb interactions can reveal interesting interaction-related effects, like energetic anomalies of the charged excitons [13] and peculiar fractional states like the $\nu = 3/8$ [14]. The optical experiments were realized using a Ti-sapphire laser with the sample mounted in a cryostat with liquid $^4$He under pumping with the magnetic field generated by a resistive magnet. The light was sent and collected through two optical fibers, and was circularly polarized using linear polarizers coupled to quarter wave plates placed in the optical paths between the fibers and the sample. In the following we will use the notation $\sigma^\pm \sigma^\pm$ to identify the polarization, indicating in successive order the polarization of the excitation and of the detection. The measurements were performed in Faraday configuration with nearly backscattering geometry, being the angle of incidence and collection of only a few degrees.

Fig. 1 shows the energy of the PL peaks as a function of the magnetic field ($B$). Several transitions can be observed, due to recombination of the excitons formed by photo-generated holes and electrons of the 2DEG merged

2with the photo-generated electrons. The labels $L_0$, $L_1$ and $L_2$ identify the transition between electrons and holes in their first, second and third Landau level (LL), respectively. We can see that at 4.1 T the transitions from the second electron LL, *i.e.* $L_1$, disappear. This is due to the empting of this LL which takes place for $\nu > 2$ since the Fermi level drops to the first LL, and allows us to identify the position of $\nu = 2$.

The lower energy transition observed in the two co-polarized spectra comes from electrons in the Zeeman split lower-lying LLs, namely $L_0^-$ (lying at the lower energy, for transition between +1/2 spin electrons and -3/2 spin holes) and $L_0^+$ (lying at higher energy, due to transitions between -1/2 spin electrons and +3/2 spin holes). In the insets of Fig. 1 we show the PL intensity as a function of $B$ for the $\sigma^+\sigma^+$ and $\sigma^-\sigma^-$ co-polarized spectra. The PL intensity has a clear reduction at $\nu = 1$ (at $B$ = 8.2 T) for both $\sigma^+\sigma^+$ and $\sigma^-\sigma^-$ polarizations. This behavior is attributed to a strong inhibition of the radiative recombination related to localization and changes in the screening response of the electron gas, which leads to reduced electron-hole overlap and broadening of the density of state [15, 16].

In Fig. 2 we present the RRS response of the quantum Hall ferromagnet around $\nu = 1$. The figure shows the series of spectra of the collected light, which include either the PL and the scattered light, measured at $\nu = 1$ for the two $\sigma^+\sigma^+$ (a) and $\sigma^-\sigma^-$ (b) polarizations. These spectra where recorded with a CCD detector, scanning continuously the energy of the laser light in the energy range of the $L_0^+$ and $L_0^-$ levels. We can observe that the $\sigma^+\sigma^+$ polarized spectra show a clear RRS at the same energy of the PL emission from $L_0^+$. However the $\sigma^-\sigma^-$ spectra do not show any RRS and we found that also the measurement taken with crosses polarizations $\sigma^+\sigma^-$ and $\sigma^-\sigma^+$ do not show any enhancement in the elastic scattering, a behavior that will be explained in the following. We notice that the non resonant component of Rayleigh scattering is very high, and the extraction of the resonant component against the strong background is a demanding task.

The effect of the Coulomb screening has to be taken into account in order to describe the RRS and the photo-excited ground state close to $\nu = 1$. In one side modulation doped GaAs-AlGaAs quantum wells, the electron and the photogenerated hole usually lye in the opposite sides of the well. When the well width $d$ is greater than the magnetic length $l$, as in our case since $l \sim 9$ nm at $\nu = 1$, the electron-electron interaction dominates the electron-hole one. Consequently the electrons form a Skyrmion first which then bounds with the valence hole, forming a state called Skyrmion-hole [8] or Skyrmion exciton [7] and the RRS close to $\nu = 1$ is resonant with the PL arising from the recombination of this state.

In low dimensional systems the RRS is the resonant light scattering into all directions and is due to the localized excitonic states [11, 12]. In our case around $\nu = 1$ the quantum Hall localization of the Skyrmion excitons adds up to the localization due to the unintentional fluctuations of the quantum well interfaces and to the one due to the finite exciton lifetime and diffusion length. The quantitative evaluation of the RRS observed in our system is complex, the origin of this resonant scattering has been for instance studied using Anderson models [11, 12] and solutions of the scattering process have been proposed using low-order Feymann diagrams [17]. The simplest approximation is to assume a RRS polarization proportional to the polarization of the Skyrmion excitons and to consider full polarization at $\nu = 1$ and zero polarization when the RRS is absent. This approximation allows the comparison of the data with the simple Skyrmion model.

The quantum Hall ferromagnet at $\nu = 1$ is described in term of Skyrmion spin texture, and in the model for independent electrons proposed by Barret *et al.* [1] the spin polarization is $S_z(\nu) = S\left(\frac{2-\nu}{\nu}\right) - (S-1)$ for $\nu > 1$ and $S_z(\nu) = \frac{1}{\nu} - (2A-1)\left(\frac{1-\nu}{\nu}\right)$ for $\nu \leq 1$, where $S(A)$ is a parameter that gives the number of spin flips per unpaired quantum of flux above (below) $\nu = 1$.

In Fig. 3 we show the value of the normalized RRS scattering, as a function of $\nu$ for the four polarization configurations. These values have been taken as the ratio between the maximum RRS enhancement and the constant Rayleigh scattering intensity out of resonance. Fig. 3 reports the Skyrmion model with $S = A = 2, 3$ and 4 considering an equal value of $S$ and $A$ in order to reflects the particle-hole symmetry wich requires that the size of the Skyrmion be the same as the anti-Skyrmions [2]. We can see that the values $S = A = 2$ and 3 are the closest to the experimental points, in particular the value $S=A=3$ has been found in the past in experiments on GaAs quantum wells with similar widths [1, 2, 18], and corresponds to 3 flipped spins per quantum flux.

The presence of RRS in $\sigma^+\sigma^+$ polarization, and its absence in the $\sigma^-\sigma^-$ can be due to the correlation between Skyrmions, screening and spin waves. The energy dispersion of the RRS signal around the optical resonance has been shown to be dependent on the energy dispersion of the homogeneous linewidth and of the dephasing times [11, 12]. Theoretical calculations of Asano et al. [7] indicate that in systems with small finite values of Zeeman splitting, as in the GaAs-AlGaAs heterostructure ($g_e \sim$ -0.4), the spin and charge degree of freedom are strongly correlated, and this plays an important role in presence of a photo-excited hole. Indeed the formation of multiple spin flips aids the hole screening when the electron-electron interaction strength is comparable with the electron-hole one. One of the main results is the presence of a jump in the energy of the $\sigma^-$ polarized luminescence when $\nu$ passes across the value 1, which is absent in $\sigma^+$ polarization. Other theoretical models [8] show that at $\nu = 1$ and around, the $\sigma^-$ luminescence presents secondary emissions peaks in the low energy side of the main emission peak, which intensity and energy shift are related to the spin wave excitations [8]. The $\sigma^+$ emission is instead not modified and is a single-peaked structure. Therefore dephasing times of the homogeneous linewidth for the $\sigma^-$ emission would be affected and their energy dispersion could present rapid variation in correspondence of these secondary emissions. As a consequence the RRS would be consequently reduced, up to be no more resolvable with respect to the non-resonant scattering response.

In summary, we have used the PL spectroscopy to investigate the magneto-excitonic transitions in a high-mobility 2DEG and the RRS to probe the collective spin texture of the quantum Hall ferromagnet at $\nu = 1$. The RRS present in the $\sigma^+\sigma^+$ co-polarized measurements, reveals the Skyrmion spin texture of the 2DEG electrons in the photoexcited ground state. Instead the interplay between Skymionic effects, screening and spin wave excitations results in a modified $\sigma^-$ polarized emission, and consequently in the absence of RRS in this polarization. These findings indicate that RRS can be used as a valid tool for the study of the spin textures of the quantum Hall systems.

**ACKNOWLEDGEMENTS**

This work was supported by the projects Cariplo Foundation *QUANTDEV*, MEC FIS2009-07880, PPT310000-2009-3, JCYL SA049A10-2 and by EuroMagNET under the EU contract n. 228043.

*Present address: Department für Physik, Ludwig Maximilians Universität, 80539 München

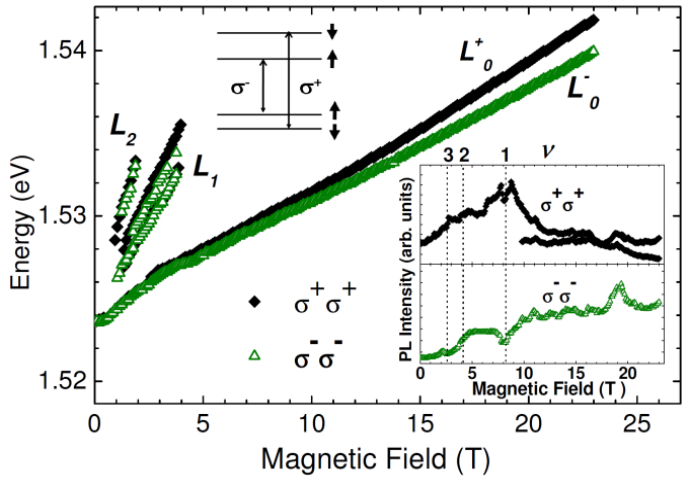

FIG. 1: Energy of the PL transitions as a function of the magnetic field at temperature of ~ 1.4 K. In the inset is reported the intensity of the PL peaks. The schematic view of the spin split level $L_0$ is also reported.

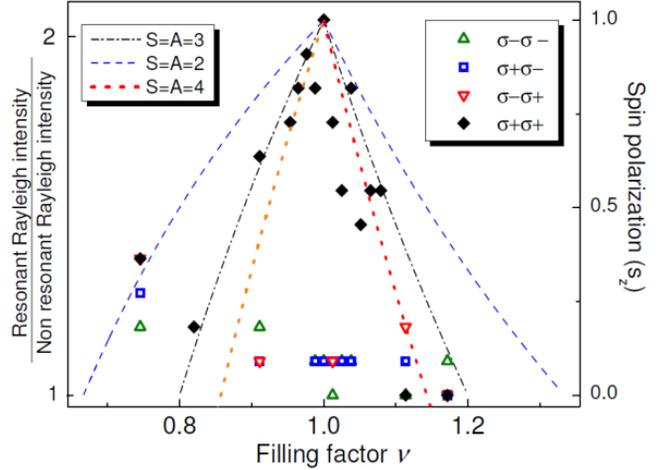

FIG. 3: The ratio between the RRS and non resonant light scattering intensities for the four crosses polarizations. The lines refers to the Skyrmion model of Ref. [1] with the different values of the parameter S=A=2, 3 and 4.

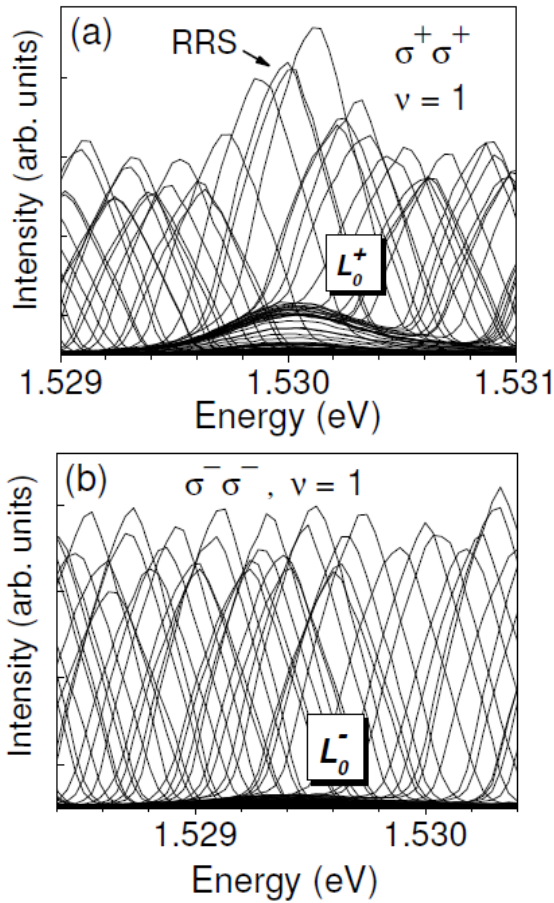

FIG. 2: The light scattering plus emission spectra around $\nu = 1$ for (a) the $\sigma^+\sigma^+$ and (b) $\sigma^-\sigma^-$ polarizations at T ~ 1.4 K.